\documentclass[pre, twocolumn, superscriptaddress, showpacs]{revtex4}
\usepackage{amsmath, amssymb, amsfonts}

\renewcommand{\[}{\begin{equation*}}
\renewcommand{\]}{\end{equation*}}

\newcommand{\elabel}[1]{\label{#1}\stepcounter{equation}\tag{\theequation}}% smart labelling...

\makeatletter
\@addtoreset{equation}{section}% reset equationnumbering each section
\makeatother
\renewcommand{\theequation}{\thesection.\arabic{equation}}% include the section's number in \theequation

\newcommand{\erw}[1]{\langle {#1} \rangle}% Expectationvalue
\def\ket#1{\mathinner{|{#1}\rangle}}% Vector
\def\bra#1{\mathinner\langle{#1}|}% Dual Vector
\def\bkt#1#2{\mathinner{\langle{#1}\,\arrowvert\,{#2}\rangle}}% Scalarproduct
\newcommand{\proj}[1]{\ket{#1}\bra{#1}}% Projecting a single variable
\newcommand{\projb}[2]{\ket{#1}\bra{#2}}% Projecting two variables
\newcommand{\CG}[6]{\left(#1#2,#3#4|#5#6\right)}% Clebsch-Gordan Coefficient

\newcommand{\mb}[1]{\text{\boldmath ${#1}$}}

\renewcommand{\Im}{\operatorname{Im}} % Imaginary Part of complex numbers
\renewcommand{\Re}{\operatorname{Re}} % Real Part of complex number 
\newcommand{\conv}[2]{\left(#1\ast#2\right)}% convolution
\newcommand{\lin}{\operatorname{lin}}% linear hull
\newcommand{\D}{{d}}% Integration
\newcommand{\order}[1]{{\mathcal O}({#1})}% Order
\newcommand{\LT}[1]{{\mathcal L}\{#1\}}% Laplacetransform
\newcommand{\LC}[2]{{\mathcal L}_{#1,#2}} % abbreviated LT of Corr. functions
\newcommand{\Id}{\textrm{Id}}% Identity
\newcommand{\R}{\mathbb{R}}% Real numbers
\newcommand{\C}{\mathbb{C}}% Complex numbers
\newcommand{\Z}{\mathbb{Z}}% Integers
\newcommand{\q}{\bm{q}}% reciprocal lattice vector
\newcommand{\mc}[1]{{\mathcal {#1}}}% mathcal symbols
\renewcommand{\th}[1]{%
\ifx1#1 \def\dummy{st}%
 \else \ifx2#1 \def\dummy{nd}%
  \else \ifx3#1 \def\dummy{rd}%
   \else \def\dummy{th}\fi\fi\fi
${#1}^{\text{\dummy}}$% this is rather dirty tricking...
}% numbering

\begin{document}
%% Front-matter
\title{Rotational Dynamics and Light-scattering in Super-cooled Molecular Liquids}
\author{Maximilian G Schultz}
%%\altaffiliation[Present address: ]{\mis}
\altaffiliation{EMail address: schultz@mis.mpg.de}
\affiliation{Freie Universit\"at Berlin, Fachbereich Physik, D-14195 Berlin, Germany}\affiliation{Max-Planck-Institut f\"ur Mathematik in den Naturwissenschaften, D-04103 Leipzig, Germany}
\author{Thomas Franosch}
\affiliation{Hahn-Meitner Institut, D-14109 Berlin, Germany}

\date{\today}
\begin{abstract}
Constitutive equations for the long-wavelength behaviour of the orientational dynamics of a super-cooled liquid are derived using a projection-operator technique and resulting expressions for light-scattering spectra are formulated. We thus extend recent studies for axially symmetric molecules to the general case of arbitrarily shaped rigid molecules. The second part of the discussion considers hydrodynamic energy-fluctuations and thus arrives at expressions for light-scattering spectra which also include a Rayleigh-line. The role of the memory-kernels in the theory is treated in detail. In particular, the derivation of a theory that formally resembles earlier approaches to the problem is presented using a mathematically rigorous description of the Laplace-transforms of correlation-functions.

\end{abstract}
\maketitle
%% Body

%% Section: Introduction {{{1
\section{Introduction}

Light-scattering methods have proven to be a powerful tool in the experimental investigation of super-cooled liquids, especially glass-forming substances, see for instance the works by Dreyfus and Pick \cite{Dreyfus98, Dreyfus99, Pick03} or those by Cummins \cite{Cummins03b}. In particular, the cross- or depolarised spectra, i.e., measurements in the VH-configuration, have revealed many details of the process of structural relaxation while cooling such liquids. As a glass-forming system approaches the glass-transition, the cross-over from pure liquid-like to solid-like behaviour is displayed in the scattering-spectra. Mode-Coupling-Theory for the glass-transition has helped to understand many of the phenomenological features of these spectra and supported the interpretation of the underlying physical processes. However, a complete theoretical formulation of the physics of light-scattering in super-cooled liquids has not been available until recently. There are theories which properly describe the Rytov-Dip in the high-temperature regime, which is due to the relaxation of shear-modes, i.e., a coupling of translational and rotational dynamics. These results stress the importance of including the rotational dynamics of the liquid in the theoretical description. On the other hand using mode-coupling-theory one can explain the appearance of the shear-waves in the dynamics at lower temperatures in place of the diffusive Rytov-Dip. The cross-over regime, however, could not be described.  Recently, several studies have been published, where phenomenologically obtained generalised Langevin equations have successfully been used to explain the light-scattering spectra of super-cooled liquids \cite{Dreyfus98, Dreyfus99, Pick03, Franosch03a}. A microscopical derivation of these equations using the long-wavelength expansion of an application of the projection-operator formalism of Mori and Zwanzig has been given for some special cases in \cite{LaLe01, Franosch03b}. Although the first attempt to fully describe the rotational dynamics using a projection-operator technique was already made by Anderson and Pecora in 1971 \cite{Anderson71}, in \cite{Franosch03b} it was shown that their theory is not suited to describe the cross-over regime properly. Anderson and Pecora did not allow for a frequency-dependence of the transport-coefficients in their theory, and their choice of variables is not suited for the application of generalised hydrodynamics \cite{Goetze89}, which is the proper framework to treat the problem and within which the present results have been derived. Our results also rely on the tensorial formalism developed by Schilling and collaborators \cite{Fabbian99, Fabbian00, Schilling97} and general results for the form of the light-scattering spectra due to symmetries \cite{FFL01}. However, the available models have simplified the physical system in question by only considering axially symmetric molecules and neglecting fluctuations of the local energy density. It is the aim of the present work to give a complete description of the rotational dynamics in the long-wavelength regime by applying the projection-operator technique of Mori and Zwanzig to the problem. The focus is mainly laid on a formal and rigorous derivation and discussion of the equations. We can therefore extend the above results to the cases of arbitrarily shaped rigid molecules and finally also include energy-fluctuations in the theory.

The paper is organised as follows. In section \ref{sec:hydro_var} we define the distinguished variables of our theory and discuss their basic properties as far as they are relevant for the application of the projection-operator formalism. Section \ref{sec:EOM} is concerned with the derivation of the generalised Langevin equations of the system in the time and the complex-frequency domain. The equations will be solved in section \ref{sec:solution}, where also the light-scattering spectra are computed and discussed. Some remarks regarding the positivity of the spectra will be made in section \ref{sec:pos_spec}. An extension of the theory that includes fluctuations of the energy-density is given in the following section. To be in accordance with earlier approaches to the problem we will finally show in section \ref{sec:Recovery} how a theory that only considers fluctuations of the dielectric tensor can be derived from ours.
%% }}}1
%% Section: Hydrodynamic Variables {{{1
\section{Hydrodynamic Variables}\label{sec:hydro_var}
We consider a classical, super-cooled liquid consisting of $N$ rigid molecules of arbitrary shape. The molecules are optically inactive and there is no orientational order present in the system. The liquid is enclosed in a volume $V$, a closed compact and simply connected subset of $\R^3$ with suitably smooth boundary, and held in equilibrium at a temperature $T$. We anticipate the thermodynamic limit $N,V \rightarrow \infty$ with the average density $n:=N/V$ being held constant. The molecules have translational and rotational degrees of freedom, which are centre-of-mass positions $\mb{R}_\alpha$ and Euler-angles $\Omega_\alpha$ respectively. The Euler-angles describe the relative orientation of the principal axes of the tensor of inertia or the polarisability tensor with respect to the laboratory frame of reference. The two interpretations only differ by a global rotation which the system is invariant under. The phase-space $\bm{\Gamma}$ is identified with $\R^{6N+6N}$. The dynamics of the system is driven by the its Poisson-brackets or Liouville-operator respectively, $\dot{A} = \{H,A\} =: i\mc{L}A$, where $H$ denotes the Hamilton-function of the liquid, 
\[
H := \sum_{\alpha=1}^{N}\biggl[\frac{\mb{P}_\alpha^2}{2m} + \sum_{i=1}^{3}\frac{J_{\alpha,i}^2}{2I_i}\biggr] + V(\{\mb{R}_\alpha,\Omega_\alpha\}).
\]
The symbol $\mb{P}_\alpha$ denotes the $\alpha$th particle's translational momentum, $J_{\alpha,i}$ labels the $i$th component of the angular momentum of particle $\alpha$ and $I_i$ is the respective eigenvalue of the tensor of inertia. The relation between the moments of inertia and the canonical rotational momentum is given in \cite{GG84}. Thus the volume-element of phase-space reads 
\[
\D\Gamma = \prod_\alpha\D\mb{R}_\alpha\D\mb{P}_\alpha\D\Omega_\alpha\D\mb{J}_{\alpha}.
\]
We assume the interaction energy $V$ only to depend on positions $\mb{R}_\alpha$ and Euler-angles $\Omega_\alpha$, but neither on translational nor orientational momenta. As a phase-space variable we understand a function $A: \mb{\Gamma}\rightarrow U$ where $U$ is a finite dimensional real or complex vector-space such that $A$ is an element of $L^1_\rho(\bm{\Gamma})\cap L^2_\rho(\bm{\Gamma})$. The subscript $\rho$ indicates integration with respect to the equilibrium measure $\exp(-\beta H)Z^{-1}$, i.e.,~thermal averaging $\erw{\cdot}$. Phase-space variables may also depend on an additional parameter, say $\bm{r} \in V$, and are assumed to have a Fourier-transform in this parameter. We will only consider fluctuations of phase-space variables, $\delta A:=A-\erw{A}$, where in the following the $\delta$ is implied by an abuse of notation. Statistical correlations are written in terms of the Kubo-scalar-product $\bkt{A}{B}:=\erw{\delta A^\ast \delta B}$. With this construction all integrals are well-defined and the space of phase-space variables becomes a Hilbert-space. Parameter-dependent variables will further be represented by their corresponding spatial Fourier-modes $A(\q):=\int_{V}e^{i\q\cdot\mb{r}}A(\mb{r})\, \D\mb{r}$. Since we are free to define the orientation of the laboratory frame we will choose $\q$ parallel to the system's z-axis. The variables we assume to exhibit hydrodynamical behaviour and will therefore be treated using a projection-operator approach comprise the fluctuations of the mass-density
\[\elabel{eq:def_rho}
\rho(\q) := \frac{m}{\sqrt{N}}\sum_{\alpha=1}^{N} e^{i\q\bm{R}_\alpha}
\]
and the mass-current,
\[\elabel{eq:def_J}
\bm{J}(\q) := \frac{1}{\sqrt{N}}\sum_{\alpha=1}^{N} \bm{P}_\alpha e^{i\q\bm{R}_\alpha} = mn\bm{v}(\q),
\]
and variables describing collective rotational motion. By definition, the Hamilton-function is invariant under spatial rotations which favours a description relying on irreducible representations of the rotation-group $SO(3)$, i.e.~so-called spherical coordinates. For each solution of the equations of motion its whole orbit under the group action also is a solution. Thus every solution carries a representation of the group. The finite irreducible representations of $SO(3)$ are unitary linear maps $\cal D$ which are classified by rank $j$. The helicity $\nu\in [-j,j]\,\cap\Z$ denotes the $\nu$th component of each vector in such a subspace. 

In this notion the mass-density is a scalar variable ($j=0$), and the mass-current a vector ($j=1$). 
To describe the orientation of a molecule and the quantities involved in light-scattering, namely the particle's polarisability tensor \cite{BP00}, we will use a $j=2$ object, i.e., a symmetric traceless tensor of rank two. 
We consider a single molecule's polarisability tensor $Q^B_{m}$, a $j=2$ object with helicity $m$, which is given in a certain body-fixed coordinate-system, rotate it back into the laboratory frame of reference and add up the contribution for the whole liquid. Thus we find as a variable describing the relevant orientational properties
\[\elabel{eq:def_Q}
Q_{\mu}(\q) := \frac{1}{\sqrt{N}}\sum_{\alpha=1}^{N} e^{i\q\bm{R}_\alpha} {\cal D}^{(2)}_{\mu m}(\Omega_\alpha) Q^B_{m},
\]
where the quantities ${\cal D}^{(2)}_{\mu m}(\Omega)$ are irreducible representations of the rotation-group of rank $j=2$, rotation-matrices on $\R^{5}$. 
(NB: We adopt the summation convention for all latin-typeset indices.) This form of the variable $Q_\mu$ and the fact that all particles have the same local properties suggest the use of rotation-matrices $\mc{D}^{(2)}_{\mu m}(\Omega_\alpha)$ as the primary objects of interest. The definition of those follows \cite{GG84}. To be in accordance with other theoretical work in the field, e.g.~\cite{Fabbian00, LaLe01}, we define the orientational density
\[\elabel{eq:def_rho_mn}
\rho_{\mu m}(\q) := i^2\sqrt{\frac{5}{N}}\sum_{\alpha}e^{i\q\bm{R}_\alpha}{\cal D}^{(2)}_{\mu m}(\Omega_\alpha).
\]
In the small $\q$-limit the time-derivative of this density will also be considered hydrodynamic. This assumption is important for proper modelling of the memory-kernels as was already pointed out in \cite{Franosch03b}. We define 
\[\elabel{eq:EoC_oricurr}
j_{\mu m}:= \dot{\rho}_{\mu m}
\] 
and call it an orientational current, although it is not derived from a conservation law. Finally, we will also consider long-wavelength energy-fluctuations. But to keep consistent with \cite{Franosch03a, Franosch03b} we will neglect them for the time being, which is justified due to the closeness of the Landau-Plazcek ratio to unity. The necessary extensions to the theory will be presented in section \ref{sec:energy}.
A treatment of general tensors of arbitrary rank is also possible. Extensive formal work has been done by Schilling {\em et~al.} \cite{Fabbian00, Schilling02}.

%% Subsection: Static Averages {{{2
\subsection{Static Averages}\label{sec:stat_avg}
Static correlation-functions, which determine orthogonality and normalisation of the dynamical variables, can be evaluated using the rotational invariance of the Hamiltonian function. According to the selection rules presented in Appendix \ref{app:norm} the $\q=0$ averages of two functions are diagonal with respect to both helicity and rank. For finite $\q$ one may expand the correlator in terms of the wave-vector about the value $\q=0$. According to the selection-rules of Appendix \ref{app:sel_rule} fixing the wave-vector parallel to $\bm{e}_z$, i.e., in spherical tensor notation $\q=q\bm{e}_0$ achieves decoupling of phase-space variables with respect to helicity. Then the mass-density-orientation correlator has the first non-vanishing term at order $\q^2$, cf.~(\ref{eq:sel_02}). In the long-wavelength limit mass-density and orientational density may therefore be considered orthogonal. Also, we can incorporate time-inversion-parity and establish orthogonality between densities and currents. 

The correlator between the orientational currents and the mass-current $\bkt{\dot{\rho}_{\mu m}}{J_{\mu}}$ is identically zero. 
This is proven easily. The orientational current breaks up into a translational part, which is linear in $\mb{P}$, and a rotational part which contains the derivatives of the Euler-angles. The interaction energy in the Hamilton-function is independent of the momenta. Therefore, the correlation of the translational part of $j_{\mu m}$ with the mass-current involves taking the expectation-value of the Wigner-functions themselves which by the assumptions on the system vanishes. The correlation of the rotational part of the orientational current with the mass-current vanishes due to gaussian averaging over the translational momenta in the mass-current. Hence in the long-wavelength limit the hydrodynamical variables mass-density, mass-current, orientational density, and orientational current are mutually orthogonal. 

Their normalisations read, evaluated at $\q=0$,
\[\elabel{eq:def_norm}
S_{mn} := \bkt{\rho_{\mu m}}{\rho_{\mu n}} \quad\text{and}\quad \Omega_{mn} := \bkt{\dot{\rho}_{\mu m}}{\dot{\rho}_{\mu n}}.
\]
In Appendix \ref{app:norm} we will show that these matrices are positive definite, and hence invertible.
Using the static structure factor $S(\q\rightarrow 0)$ and the associated isothermal sound-velocity $Sc^2:=\frac{k_BT}{m}=:v^2$ we have
\[
\bkt{\rho}{\rho} = \frac{m^2v^2}{c^2} \quad\text{and}\quad  \bkt{J_\mu(\q)}{J_\nu(\q)} = \delta_{\mu\nu}m^2v^2.
\]
%% }}}2
%% Subsection: Equations of Continuity {{{2
\subsection{Equations of Continuity}
Both the mass-density and its current obey equations of continuity. The conservation law for the mass-density in spherical coordinates is the same as in cartesian coordinates
\[\elabel{eq:EoC_mass}
\dot{\rho}(\q,t) = i q_{-m}J_m(\q,t).
\]
The conservation of momentum is achieved by using the stress tensor $\Pi$ which is a tensor having only its $j=0$ and $j=2$ components non-vanishing, because the cartesian representation of $\Pi$ is a symmetric matrix. The conservation law can be derived from the cartesian version, cf. for instance \cite{Franosch03b}, using the known transformation laws \cite{GG84}, 
\[\elabel{eq:EoC_curr}
\dot{J}_{\nu}(\q) = iq \sum_{j}\CG{1}{\nu}{1}{0}{j}{\nu}\,\Pi^{(j)}_\nu(\q).
\]
The coefficients of the linear combination in (\ref{eq:EoC_curr}) are Clebsch-Gordan-coefficients. Also, the relation $\q=q\bm{e}_0$ has been used here. Therefore the sum contains only terms for $j=0,2$. The scalar part, i.e., the $j=0$ component, is interpreted as pressure, whereas the tensorial part, i.e., the $j=2$ components, describe shear.
%% }}}2
%% }}}1
%% Section: Equations of Motion {{{
\section{The Projection-Operator Technique}\label{sec:EOM}

In this section, we will derive the equations of motion for the variables introduced before by applying the projection-operator technique of Zwanzig and Mori \cite{Forster75, Hansen90}. The thus obtained equations have the form of generalised Langevin equations. The coupling between the conserved quantities and the orientational degrees of freedom will be achieved through the use of frequency-dependent memory-kernels which will have the function of transport coefficients. A light-scattering experiment only probes the dynamics of the liquid on a spatial scale which is large compared to any microscopic length-scale. It is therefore sufficient to derive the equations in leading order of the wave-vector $\q$. Although the spatial scales are well separated the time-scales are not. It will turn out that by using frequency dependent memory-kernels the theory is able to describe the dynamics on a wide range of scales they as can be observed experimentally. 

%% Subsection: Projection Operator {{{2
\subsection{Projection Operator}
In order to close the system of equations (\ref{eq:EoC_oricurr}, \ref{eq:EoC_mass}, \ref{eq:EoC_curr}) we need to derive constitutive equations for $\Pi^{(0)}$, $\Pi^{(2)}$, and the orientational force $\ddot{\rho}_{\mu n}$. 
As pointed out before we will apply a projection-operator technique. Then the dynamics will naturally be split into three parts, the intrinsic dynamics of the hydrodynamic subspace, the mixing dynamics and those in the orthogonal complement. The latter will generally be interpreted as the 'bath degrees of freedom', because their correlation to the distinguished variables vanish. The variables onto which we will project the dynamics have already been introduced in section \ref{sec:hydro_var}, leading to the following projection-operator 
\[\elabel{eq:Project}
\begin{split}
{\cal P}(\q) := &\sum_{\mu}(S^{-1})_{mn}\projb{\rho_{\mu m}(\q)}{\rho_{\mu n}(\q)}\\& + \sum_{\mu}(\Omega^{-1})_{mn}\projb{\dot{\rho}_{\mu m}(\q)}{\dot{\rho}_{\mu n}(\q)}\\
&+ \proj{\rho(\q)}\frac{c^2}{m^2v^2}\\& + \sum_{\mu}\frac{\proj{J_\mu(\q)}}{m^2v^2}.
\end{split}
\]
We note that $\cal P$ is a projection only up to order $\order{\q^2}$, which is the convenient setting for the long-wavelength studies discussed in this paper. The time-evolution generated by the Liouville-operator is given by an operator $R(t)$ which can be rewritten using the projection-operators $\mc{P}$ and $\mc{Q}:=1-\mc{P}$, \cite{Franosch03b}
\[\elabel{eq:semigroup}
R(t) = R(t){\cal P} + \int_{0}^{t}R(s){\cal P}i{\cal L}R'(t-s) \D s + R'(t).
\]
A reduced time-evolution-operator has thereby been defined by
\[
R'(t):=\mc{Q}e^{i\mc{QLQ}t}\mc{Q}.
\]
Equation (\ref{eq:semigroup}) is completely equivalent to the generalised Langevin equations discussed in \cite{Hansen90, Forster75}. For each phase-space-variable $A$ the evolved variable $R'A$ is orthogonal to each of the hydrodynamic variables. This property identifies it as {\em noise} and justifies its neglect for the remainder of this paper.\\
We can simplify the integral-kernel $\mc{P}i\mc{L}R'$ by operating with $\mc{L}$ to the left,
\[\elabel{eq:semigroup_PiL}
\begin{split}
& {\cal P}(-i{\cal L})R' = 
\sum_{\mu}(\Omega^{-1})_{mn}\projb{\dot{\rho}_{\mu m}(\q)}{\ddot{\rho}_{\mu n}(\q)}R'\\
&- \sum_{\mu}\frac{iq}{m^2v^2}\projb{J_\mu(\q)}{\Pi^{(j)}_{\mu}(\q)}\CG{1}{\mu}{1}{0}{j}{\mu}R'.
\end{split}
\]
The application of the evolution-operator (\ref{eq:semigroup}) to stress tensor and orientational force yields the missing equations of motion. However, the price to be paid is the introduction of frequency dependent memory-kernels. Because the reduced dynamics do not show any hydrodynamic poles, these kernels can be approximated by their $\q=0$ value.

A light-scattering experiment measures the Laplace-transform of the autocorrelation of the dielectric tensor or more precisely the real-frequency-limit of its imaginary part \cite{BP00}. A formulation of the equations of motion to directly yield these correlator is more useful than the formulation in terms of generalised Langevin equations. Also, the integro-differential equations will reduce to matrix-equations which can easily be solved. However, the equations in the time-domain cannot be derived from the matrix counterparts as both the Laplace-transform and canonical averaging are not invertible. We will therefore perform all calculations in the complex-frequency domain, but also present the equations of motion in the time-domain, where the individual couplings between the various physical quantities can readily be read off.
%% }}}2
%% Subsection: The Correlation Matrix {{{2
\subsection{The Correlation Matrix}

According to \cite{Forster75} the Laplace-transform of a correlation matrix $C(z)$ of a vector of hydrodynamic variables $A=(A_1,\dots,A_n)$ obeys the following equation
\[\elabel{eq:MZW_2}
C(z) \cdot S^{-1} \left[ Sz -i \bkt{A}{\dot{A}} + \bkt{\dot{A}}{R'\dot{A}}\right] \cdot S^{-1}= -\Id,
\]
where $\Id$ denotes the unit-matrix. All variables appear in their Laplace-transformed form which is conventionally set to
$\LT{f(t)}(z) := i\int_{0}^{\infty}e^{izt}f(t)\,\D t$.
The symbol $S$ denotes the static correlation-matrix $S_{ij}:=\bkt{A_i}{A_j}$. Given this formula we only need to evaluate a few static and dynamic correlation-functions to set up the equations of motion. Due to time-inversion parities and rotational invariance of the Hamilton-function the only non-vanishing static correlators are
\[\elabel{eq:stat_corr}
\bkt{\rho_{\mu m}}{\ddot{\rho}_{\mu n}} = -\Omega_{mn}
\quad\text{and}
\quad\bkt{\rho}{\dot{J}_{0}} = -m^2v^2.
\]
The dynamical correlation-functions involving the reduced time-evolution operator $R'(t)$ are memory-kernels which can be related to transport-coefficients using the so-called Green-Kubo relations \cite{Forster75}. When calculating the light-scattering spectra we can identify the transport-coefficients by comparing with the theory of the simple fluid. Most of these coefficients are of order $\order{q^2}$ and vanish in the long-wavelength limit. If we set $A_i$ to one of those densities whose current is an element of the hydrodynamic subspace the reduced dynamical correlator will vanish, because then $\mc{Q}\dot{A}=0$. This applies to all correlators involving $\dot{\rho}$ and $\dot{\rho}_{\mu m}$. Setting $A_i=J_\mu$ the correlator can be rephrased using the equation of continuity (\ref{eq:EoC_curr}) and results in the autocorrelations of pressure $p :=\frac{-1}{\sqrt{3}} \Pi^{(0)}$ and the shears $\Pi^{(2)}$. For the non-vanishing elements of the matrix $\bkt{\dot{A}}{R'\dot{A}}$ we find
\[
\bkt{p(\q)}{R'(z)p(\q)}=: \tilde{\eta}_b(z) + \order{q^2}
\]
defining a bulk viscosity $\eta_b$. The correlation between orientational forces and pressure defines translation-rotation couplings
\[
\bkt{\ddot{\rho}_{\mu n}}{R'(z)\Pi^{(2)}_\nu} =: -\delta_{\mu\nu}\tilde{\mu}_n(z) + \order{q^2}.
\]
The autocorrelation of shear reads
\[
\bkt{\Pi^{(2)}_\mu(\q)}{R'(z)\Pi^{(2)}_{\mu}(\q)} =: \tilde{\eta}_s(z) + \order{q^2}.
\]
Finally, the reduced autocorrelation of the orientational forces defines a rotational friction
\[
\bkt{\ddot{\rho}_{\mu n}(\q)}{R'(z)\ddot{\rho}_{\mu r}(\q)} =: \tilde{\Gamma}_{nr}(z). 
\]
With these definitions and the decoupling of correlators with respect to helicities (\ref{eq:sel_rule_gen}) we find three matrices representing the operator $\tilde{M}(z) := Sz -i \bkt{A}{\dot{A}} + \bkt{\dot{A}}{R'\dot{A}}$, one for each modulus of helicity. For notational conveniences we define some shorthands 
\[\elabel{eq:def_eta}
C_\nu:=\CG{1}{\nu}{1}{0}{2}{\nu}\quad\text{and}\quad 
\tilde{\eta}:= C_0^2\tilde{\eta}_s + \tilde{\eta}_b.
\]
Using block-matrix notation and $A:=(\rho_{\nu r}, \dot{\rho}_{\nu r}, J_\nu, \rho)$ where $r \in [-\nu;\nu]\cap \Z$ we arrive at
\[\elabel{eq:M_pm2}
\tilde{M}_{\nu=\pm2}=
\begin{pmatrix}
zS_{nr} & i\Omega_{nr} \\
-i\Omega_{nr} & \left( z\Omega_{nr} + \tilde{\Gamma}_{nr}\right)
\end{pmatrix}
\]

\[\elabel{eq:M_pm1}
\tilde{M}_{\nu=\pm1} = \begin{pmatrix}
zS_{nr} & i\Omega_{nr} & 0 \\
-i\Omega_{nr} & z\Omega_{nr} + \tilde{\Gamma}_{nr} & -iqC_{\pm1}\tilde{\mu}_n\\
0 & iqC_{\pm1}\tilde{\mu}_r^T  & C_{\pm1}^2q^2\tilde{\eta}_s + m^2v^2z
\end{pmatrix}
\]

\[\elabel{eq:M_0}
\tilde{M}_{\nu=0} = \begin{pmatrix}
zS_{nr} & i\Omega_{nr} & 0 & 0 \\
-i\Omega_{nr} & z\Omega_{nr} + \tilde{\Gamma}_{nr} & -iqC_0\tilde{\mu}_n & 0\\
0 & iqC_0\tilde{\mu}_r^T & q^2\tilde{\eta} + zm^2v^2 & qm^2v^2\\
0 & 0 & qm^2v^2 & z \frac{m^2v^2}{c^2} 
\end{pmatrix}.
\]
In this form, the equations of motion (\ref{eq:MZW_2}) can be solved and the scattering spectra written down in a microscopically exact form. There is an alternative approach going the route via constitutive equations in the time domain. The explicit procedure has been demonstrated eg.~in \cite{Franosch03b}. Here, we will just give the results for our set of variables without explicitly mentioning the noise term. A couple of abbreviations are introduced. First we absorb the frequency and structure-factor matrices into the corresponding memory-kernels, 
\begin{gather}
\mu_m(\tau):=(\Omega^{-1})_{mn}\tilde{\mu}_n(\tau),\\
\Gamma_{mr}(\tau) := (\Omega^{-1})_{mn}\tilde{\Gamma}_{nr}(\tau),
\end{gather}
and define
\begin{gather}
\omega_{mr}:= (S^{-1})_{mn}(\Omega)_{nr},\\
(\Lambda')_{mr}:=(\Omega)_{mr}\frac{n}{k_BT}.
\end{gather}
The viscosities with and without tilde are related to each other by a factor of $\frac{n}{k_BT}$.
%% Since we have put $\q$ parallel to the z-axis, the helicity zero equation for the mass-current is trivial, the equation of continuity. The nontrivial helicities $\nu=\pm1$ obey
The equations for the mass-current turn out to be
\begin{multline}
\dot{J}_{\pm1}(\q,t) = iqC_{\pm1}\conv{\mu_m}{\dot{\rho}_{\pm\!1,m}(\q)}(t)\\
- \frac{q^2}{mn}C_{\pm1}^2\conv{\eta_s}{J_{\pm1}(\q)}(t)
\end{multline}
and
\begin{multline}
\dot{J}_0(\q,t)  =  iq\Bigl[c^2\rho + \frac{iq}{mn}\conv{\eta_b}{J_0(\q)}(t)\Bigr]\\ + iqC_0\Bigl[\conv{\mu_m}{\dot{\rho}^{(2)}_{0 m}(\q)}(t) + \frac{iqC_0}{mn}\conv{\eta_s}{J_0(\q)}(t)\Bigr],
\end{multline}
where the asterisk denotes convolution in time. Therefore, using (\ref{eq:EoC_mass}), the equation for the mass-density is
\begin{multline}
\ddot{\rho}(\q,t) = -q^2c^2\rho(\q,t) - \frac{q^2}{mn}\conv{\eta}{\dot{\rho}(q)}(t)\\
 - q^2 C_{0} \conv{\mu_m}{\dot{\rho}_{0,m}(q)}(s).
\end{multline}
The equations for the orientational densities, with $\nu=0,\pm1$ are
\begin{multline}
\ddot{\rho}_{\nu r}(q,t) = - \omega_{mr}\rho_{\nu m}(q,t)%
 -\conv{\Gamma_{mr}}{\dot{\rho}_{\nu m}(q)}(s)\\%
- (\Lambda')_{mr} C_{\nu}\frac{iq}{mn}\conv{\mu_m}{J_\nu(q)}(s).\label{eq:rho0pm1}
\end{multline}
And for $\nu=\pm2$ one finds
\[\elabel{eq:rhopm2}
\ddot{\rho}_{\pm2,r}(q,t) = - \omega_{mr}\rho_{\pm2, m}(q,t) - \conv{\Gamma_{mr}}{\dot{\rho}_{\pm2, m}(q)}(s).
\]
These equations have the same functional form as the ones presented in \cite{Franosch03b}. Due to the use of Wigner-functions instead of a single orientational variable they form a vector-valued theory in contrast to the scalar theory developed in \cite{Franosch03b}. For a comparison of the two approaches see section \ref{sec:Recovery}.
%% }}}2
%% }}}1
%% Section: Solution of the Dynamical Equations {{{1
\section{Solution of the Dynamical Equations, Light-scattering Spectra}\label{sec:solution}
Due to the formulation of the problem in terms of Laplace-transforms of correlation-functions, we only have to invert matrices in order to solve (\ref{eq:MZW_2}). The presentation of the results proceeds in three steps. First we solve the equations for helicities $\nu=\pm2$, then $\nu=\pm1$, and finally the $\nu=0$ part. This allows to go from the most simple coupling, namely none at all, towards the coupling of all variables. For notational convenience we abbreviate the Laplace-transforms of the dynamical variables by ${\cal L}_{sr}^{\nu} := \LT{\bkt{\rho_{\nu s}}{\rho_{\nu,r}}}$ where $\nu$ is defined by the context the equation is given and mostly omitted at all, 
and $\mc{L}_{\rho\rho} :=\LT{\bkt{\rho}{\rho}}$. 

The helicities $\nu=\pm2$ only concern orientational variables,
\[\elabel{eq:sol_pm2}
\mc{L}_{sm}^{\pm2}(z) = S_{sn}\left(\delta_{nr}z + \Gamma_{nr}(z)\right)D^{-1}_{rm}(z).
\]
where we have defined 
\[\elabel{eq:D_mr}
D_{mr}(z) :=-z^2 \delta_{mr} + \omega_{mr} -z \Gamma_{mr}(z).
\]
Due to the additional coupling of orientation to the mass current for $\nu=\pm1$ one will find more diverse dynamics in these variables. Defining a damping $\check{\eta}(z)$ by
\[\elabel{eq:damp_eta}
\check{\eta}(z) := \eta_s(z) + z\mu_n(z)(\Lambda')_{nr}D^{-1}_{rm}(z)\mu_m(z).
\]
the solution for the orientational variables reads
\[
\elabel{eq:sol_pm1}
\mc{L}_{sm}^{\pm1}(z) = \mc{L}_{sm}^{\pm2}(z) - q^2\frac{C_{\pm1}^2\alpha_m(z)\alpha_s(z)m^2v^2}{z + \frac{q^2}{mn}C_{\pm1}^2\check{\eta}(z)}.
\]
where an amplitude
\[\elabel{eq:def_alpha}
\alpha_m(z):=\frac{1}{v^2}\mu_n(z)\Omega_{nr}D^{-1}_{rm}(z)
\]
has been introduced. The $\nu=0$ components exhibit the most structure, for there are also couplings to the fluctuations of the mass-density.
Introducing another damping-kernel
\[
\elabel{eq:damp_gamma}
K(z) := \eta_b(z) + C_{0}^2\check{\eta}(z).
\]
one finds the density-density auto-correlator
\[\elabel{eq:sol_dens}
\mc{L}_{\rho\rho}(z) = \frac{m^2v^2}{c^2z}\left[ \frac{q^2c^2}{q^2c^2-z^2-\frac{q^2}{mn}zK(z)}-1\right].
\]
The autocorrelation-function of the orientational variables is 
\[
\elabel{sol_0}
\mc{L}_{sm}^{0}(z) = \mc{L}_{sm}^{\pm2}(z)%
+ q^2\frac{C_{0}^2\alpha_m(z)\alpha_s(z)zm^2v^2}{q^2c^2-z^2 - \frac{q^2}{mn}zK(z)}.
\]
And finally the correlation of density and orientation reads
\[\elabel{eq:sol_rho_ori}
\mc{L}_{\rho,0}^0(z) = C_0q^2\alpha_m(z)\frac{m^2v^2}{q^2c^2 -z^2 -\frac{q^2}{mn}zK(z)}.
\]
%% }}}2
%% Subsection: Spectra and Analysis {{{2
\subsection{Light-Scattering Spectra}
In order to describe a light-scattering experiment a model for the dielectric-tensor has to be assumed. Here, we stick to the model of \cite{Franosch03a, Franosch03b} where $\delta\varepsilon:=a\rho + Q$ is the sum of fluctuations of the mass-density and contributions of the orientational dynamics. The general dielectric tensor is a $3\times 3$ matrix, thus a nine-dimensional object. Its representation in terms of irreducible representations of the rotation group is lying in a direct sum of the invariant subspaces with $j=0, 1, 2$. The molecules of the liquid have been assumed to be optically inactive. This yields a symmetric dielectric tensor, thus having the $j=1$ component vanishing. As discussed in section \ref{sec:hydro_var} the mass-density fluctuations will account for the scalar part, whereas the orientational variables, defined in (\ref{eq:def_Q}), give the tensorial component. The exact derivation of the scattering-spectra in cartesian coordinates can be found in \cite{BP00}. Using the transformation laws between cartesian and spherical representations \cite{GG84} one obtains up to an overall multiplicative constant the following expressions.
The polarised spectrum is given by
\[\elabel{eq:spectra1}
I^{VV} = \Biggl\{ \frac{1}{3}{\cal L}_{0,0}^0 + \frac{\sqrt{2}}{3}{\cal L}_{0,2}^0 + \frac{1}{6}{\cal L}_{2,2}^0 + \frac{1}{2} {\cal L}_{2,2}^2\Biggr\}.
\]
And the cross-polarised (in the literature this is also noted as `depolarised') spectra reads
\[\elabel{eq:spectra2}
I^{VH} = \frac{1}{2}\Biggl\{ \sin^2\!\left(\frac{\vartheta}{2}\right){\cal L}_{2,2}^2 + \cos^2\!\left(\frac{\vartheta}{2}\right){\cal L}_{2,2}^1\Biggr\}.
\]
The dielectric tensor has been identified as an object $\delta\varepsilon^{j}_\nu$ where $j=0,2$. The according Laplace-transforms of its autocorrelation-functions and the limit of the imaginary part to the real axis, which is the actually measured quantity, have been abbreviated 
\[
{\cal L}_{j,k}^{\nu}(\omega) := \lim_{\varepsilon \searrow 0}\Im\LT{\bkt{\varepsilon^j_\nu(t)}{\varepsilon^k_\nu(0)}}(\omega+i\varepsilon),
\]
where $\omega \in \R$ and $\varepsilon>0$. The scattering-angle is denoted $\vartheta$. The correlators of the orientational components are obtained by multiplying those of the orientational densities $\rho_{\mu m}$ by the local polarisability tensor $Q^B_m$ which without loss of generality we can assume to be diagonal. Plugging in the expressions obtained in the previous section one finds the following
\begin{widetext}
\[\elabel{eq:I_VV_final}
I^{VV} = \frac{1}{3}\lim_{\varepsilon\searrow 0}\Im\Biggl\{ \frac{4}{10z}\mb{Q}^{B}\cdot\mb{\Omega}\mb{D}^{-1}\cdot\mb{Q}^{B}%
+q^2\frac{1}{z}\frac{m^2v^2}{q^2c^2-z^2-\frac{q^2}{mn}zK(z)}\biggl[ a -\frac{z}{\sqrt{10}}C_0\mb{\alpha}\cdot\mb{Q}^{B}\biggr]^2\Biggl\}
\]
\[\elabel{eq:I_VH_final}
I^{VH} =  \frac{1}{10}\lim_{\varepsilon\searrow 0}\Im\mb{Q}^{B}\Biggl\{ \frac{1}{z}\mb{\Omega}\mb{D}^{-1} 
- \cos^2\Bigl(\frac{\vartheta}{2}\Bigr)\,q^2C_{\pm1}^2\frac{m^2v^2}{z + \frac{q^2}{mn}C_{\pm1}^2\check{\eta}(z)}\mb{\alpha}(z)\,\mb{\alpha}(z)\Biggl\}\mb{Q}^{B}.
\]
\end{widetext}
NB: Bold symbols are the coordinate-free representation of the respective object, eq.~$\bm{Q} \sim Q_\mu$. 
Both spectra exhibit a purely rotational background whose amplitude is independent of $\q$. The second term of each spectrum is proportional to $\q^2$ as expected by symmetry considerations \cite{FFL01}. In (\ref{eq:I_VV_final}) this term represents a longitudinal phonon with velocity $c$, damped by the kernel $K(z)$ and having a frequency-dependent amplitude which is also determined by the translation-rotation coupling hidden in $\bm{\alpha}$. The cross-polarised spectrum on the contrary shows a primarily diffusive shear mode with the viscosity $C^2_{\pm1}\check{\eta}$. In a backscattering geometry where $\vartheta=\pi$ this term vanishes and the spectrum becomes purely rotational. Effects of structural relaxation in the orientational degrees of freedom are therefore accessible best in such a configuration. If one assumes a characteristic behaviour of the memory-kernels at high and low temperature, one can derive limiting expressions for the spectra and the occurring phenomena. At high temperatures the frequency-dependence of the transport-coefficients can be neglected and the $q^2$ mode in $I^{VH}$ therefore becomes a purely diffusive shear mode, which is subtracted off the rotational background. This is the so-called Rytov-dip \cite{Stegeman68}. At very low temperatures one can model the structural relaxation of the system by non-ergodic correlation-functions \cite{Goetze91}. The Laplace-transforms of such functions are negatively proportional to $\frac{1}{z}$. In this case the shear-mode becomes a propagating shear wave, showing the onset of solidification of the system near the glass-transition. Thus, the theory developed so far and the resulting expressions for the light-scattering spectra (\ref{eq:I_VV_final},\ref{eq:I_VH_final}) are able, along with reasonably justified approximations of the memory-kernels, to explain the cross-over in the cross-polarised spectrum from one exhibiting the Rytov-Dip to the low-temperature case, where propagating shear-waves are observed \cite{Cummins03b}. 

If one neglects that there is a contribution of the translation-rotation coupling to the longitudinal phonon dissipation kernel $K(z)$, the isotropic part of the polarised spectrum, i.e.,~that part that is independent of the anisotropy of the molecules, is the longitudinal phonon with amplitude $a$, 
\[\elabel{eq:def_I_iso}
I_{\text{iso}}:= aq^2\frac{\mc{C}^2}{3}\lim_{\varepsilon\searrow 0}\Im\frac{1}{z}\frac{m^2v^2}{q^2c^2-z^2-\frac{q^2}{mn}zK(z)}.
\]
In contrast to results of previous approaches, cf.~\cite{BP00}, where $I^{VV}=I_\text{iso}+\frac{4}{3}I^{VH}(\vartheta=\pi)$, we find an additional contribution to $I^{VV}$ due to the translation-rotation coupling, i.e.,~
\begin{widetext}
\[\elabel{eq:I_VV_decomp}
I^{VV}= I_\text{iso} + \frac{4}{3}I^{VH}(\vartheta=\pi) + q^2\frac{\mc{C}^2}{3\sqrt{5}}\lim_{\varepsilon\searrow 0}\Im\Biggl\{\frac{m^2v^2}{q^2c^2-z^2-\frac{q^2}{mn}zK(z)}\biggl[ -a\frac{2}{\sqrt{3}}\mb{\alpha}\cdot\mb{Q}^{(2),B} +\frac{z}{3}\left(\mb{\alpha}\cdot\mb{Q}^{(2),B}\right)^2\biggr]\Biggl\}.
\]
\end{widetext}
The difference $I^{VV}-(\frac{4}{3}I^{VH}(\vartheta=\pi) + I_{\text{iso}})$ has recently been measured experimentally and called the `VV-Dip' \cite{Cummins03b}. This result shows that an approach without considering the rotational dynamics or the translation-rotation-coupling is too simple and contradicted by experiments.

%% }}}2

%% }}}1
%% Section: Positivity of the Spectra {{{1
\section{Positivity of the Spectra}\label{sec:pos_spec}
An application of the theory presented in the previous sections to experiments is only possible when the memory-kernels are approximated by phenomenological functions. One problem, however, of any approximation-scheme is to find the admissible set of kernels that will lead to physical results or be able to identify those approximations which yield unphysical results. For instance models for the memory-kernels should not lead to negative scattering-spectra which renders positivity a useful criterion for the admissibility of an approximation--scheme. In this paragraph, we will discuss positivity of the correlation-matrix which solves (\ref{eq:MZW_2}) and thereby prove that the spectra will be positive as long as the memory-kernels are positive semidefinite matrices. In section \ref{sec:EOM}, we have formulated the equations of motion as matrix equations for the Laplace-transforms of the respective correlation-functions. These equations are quite simple, they all have the form
\[\elabel{eq:inv}
A\cdot B=-\Id,
\]
with $A$ and $B$ defined appropriately.
Now let $\cal A$ be the space $\C^{n\times n}$ with positive and finite integer $n$, the space of all complex $n\times n$ matrices. With the terms of real and imaginary part of a matrix $C$ we denote the following hermitian matrices, cf.~\cite{Con90},
\begin{eqnarray}
\Re C &:=& \frac{1}{2}\left(C + C^\dagger\right)\\
\Im C &:=& \frac{1}{2i}\left(C - C^\dagger\right),
\end{eqnarray}
where the dagger ${}^\dagger$ denotes transposition and complex conjugation. A hermitian matrix $A$ in $\mc{A}$ is said to be positive, i.e., $A \ge 0$, if its spectrum, which is real, also is nonnegative. Equivalently we can say it is positive semidefinite. 
Now let $A,B\,\in {\cal A}$ and $A\cdot B = -\Id$. Multiplying equation (\ref{eq:inv}) with $B^\dagger$ from the left we find $B^\dagger\cdot A \cdot B = -B^\dagger$ and the adjoint equation $B^\dagger\cdot A^\dagger \cdot B = -B$.
Now, subtracting these two one obtains
\[\elabel{eq:pos}
B^\dagger \cdot \Im A \cdot B = \Im B.
\]
It can be shown that for a positive element $A$ the transformed element $B^\dagger\cdot A \cdot B$ is still positive \cite{Con90}. Thus we know from the positivity of $\Im B$ that $\Im A$ must be a positive element of $\mc{A}$.
In the following, we will prove that the positivity of the imaginary part of a correlation matrix $C(z)$ is actually equivalent to the positivity of the imaginary part of the memory-kernel matrix $\bkt{\dot{A}}{R'\dot{A}}$ in the above sense. 

%% Subsection: Application to the Mori-Zwanzig formalism {{{2
\subsection{Application to the Mori-Zwanzig formalism}

According to equations (\ref{eq:MZW_2}, \ref{eq:pos}) we have to show that $S\tilde{M}S \ge 0$. All our considerations comprise a finite dimensional subspace of the state-space. Without loss of generality we can therefore use orthogonal variables, i.e.,~$\bkt{A_i}{A_j} = \delta_{ij}\bkt{A_i}{A_i}$
If the variables were not orthogonal we would have had to use a similarity transformation to make them orthogonal. However, the imaginary part of any matrix is hermitian, and hence the similarity transformation that diagonalises the imaginary part is unitary and will not influence any positivity properties.
In the orthogonal system, the matrix $S_{ij}$ is diagonal and, because this holds for autocorrelation-functions in general, the diagonal elements are real and positive. The same is true for its inverse $(S_{ij})^{-1}$. The multiplication by $S^{-1}$ is therefore also irrelevant for the discussion of positivity. We now analyse the three different addends in equation (\ref{eq:MZW_2}) and examine their imaginary parts.
\begin{itemize}
\item $S_{ij}$ is real and symmetric. Thus $\Im Sz = S\Im z \ge 0$ or vanishes if we write $z=\omega+i\varepsilon$ and let $\varepsilon \searrow 0$.
\item The imaginary part of the reactive contribution $-i\bkt{A_i}{\dot{A}_j}$ is identically zero, because $\bkt{A_i}{\dot{A}_j} = -\bkt{\dot{A}_i}{A_j}$.
\item The last term represents the matrix of memory-kernels which is at our disposition.
\end{itemize}
Collecting these properties and using the result about positive elements in $\C^{n\times n}$ we have proved that with the memory-kernels having positive imaginary part, the correlation-functions will also have positive imaginary part. The converse also holds due to a theorem in \cite{Akh65}.

%% }}}2
%% Subsection: Necessary condition on the memory-kernels {{{2
\subsection{Conditions on the Memory-Kernels}

We shall now apply the general result of the previous paragraph to the equations of motion of the rotational dynamics in a super-cooled liquid. We will thus find necessary and sufficient conditions for the individual memory-kernels of the theory such that the resulting memory-kernel matrix has positive imaginary part.\\
We always assume the individual memory-kernels to be proper correlation-functions, i.e.,~be positive and bounded in the sense of Appendix \ref{app:moments}.
Being given the matrices (\ref{eq:M_pm2}-\ref{eq:M_0}) we can extract their imaginary part, i.e., the imaginary part of the memory-kernel matrix $\bkt{\dot{A}}{R'\dot{A}}$. We will treat $\nu=0$ in detail here, the remainder works out by analogy,
\[\elabel{eq:mmkm}
\bkt{\dot{A}}{R'\dot{A}}_{\nu=0}=\begin{pmatrix}
0 & 0 & 0 & 0\\
0 & \tilde{\Gamma}_{nr} & -iqC_{0}\tilde{\mu}_n  & 0\\
0 & iqC_{0}\tilde{\mu}^T_r & q^2\tilde{\eta} & 0\\
0 & 0 & 0 & 0
\end{pmatrix}.
\]
Its imaginary part is the matrix
\[\elabel{eq:mmkm_im}
\begin{pmatrix}
0 & 0 & 0 & 0\\
0 & \Im\tilde{\Gamma}_{nr} & -iqC_0\Im\tilde{\mu}_n & 0\\
0 & iqC_0\Im\tilde{\mu}^T_r & q^2\Im\tilde{\eta} & 0\\
0 & 0 & 0 & 0
\end{pmatrix}.
\]
Since the matrix $\tilde{\Gamma}_{nr}$ is symmetric, the imaginary part can be taken element-wise. For the vector $\tilde{\mu}_n$ the same applies. Take a vector $\left(\mb{a}, b\right)$ where $\mb{a} \in \C^{2\nu+1}$ and $b \in \C$. Hence, positivity of (\ref{eq:mmkm_im}) is equivalent to the inequality
\[\elabel{eq:EOM_0_ineq_raw}
\begin{split}
0&\le 
\begin{pmatrix}
\mb{a} & b
\end{pmatrix}^\ast
\begin{pmatrix}
\Im\tilde{\Gamma}_{nr} & -iqC_0\Im\tilde{\mu}_n \\
iqC_0\Im\tilde{\mu}^T_m & q^2\Im\tilde{\eta}
\end{pmatrix}
\begin{pmatrix}
\mb{a}\\b
\end{pmatrix}
\\&=
\mb{a}^\ast \Im \tilde{\Gamma} \mb{a} + b^\ast (q^2\Im\tilde{\eta})b + 2\Re\left[b (iqC_0\Im\tilde{\mb{\mu}})\mb{a}\right].
\end{split}
\]
This relation holds true for all $b \in \C$ and in particular for the minimising value $b_{\text{min}}$. Due to the convexity of $b^\ast(q^2\Im \tilde{\eta})b$ in $b$, the global minimum is the only critical value of the right-hand-side. Differentiating the above inequality with respect to $b$ and equating to zero yields
\[
b_{\text{min}}^\ast=-\frac{iqC_0\Im(\tilde{\mb{\mu}})\mb{a}}{q^2\Im\tilde{\eta}}.
\]
Inserting into (\ref{eq:EOM_0_ineq_raw}) we find that for positivity of the operator in question the memory-kernels need to fulfil
\[\elabel{eq:EOM_0_ineq}
\Im\tilde{\eta}(\mb{a}^\ast\Im\tilde\Gamma\mb{a}) - |\Im(\tilde{\mb{\mu}})\cdot\mb{a}|^2 \ge 0,
\]
which is both necessary and sufficient for the assertion to hold true. Condition (\ref{eq:EOM_0_ineq}) resembles the one found in \cite{Franosch03b}. However, this result as the application of the positivity properties of Mori-Zwanzig-projections is much more general as it is derived using a method that does not rely on the specific properties of the system in question.
%% }}}2

%% }}}1
%% Section: Including Temperature {{{1
\section{Energy Fluctuations}\label{sec:energy}
In the preceding discussion, we have neglected temperature, i.e.,~energy-density, as a hydrodynamic variable. A complete theory of super-cooled molecular liquids should include the hydrodynamic mode which is due to the conservation of energy. In the available literature, fluctuations of the energy-density are usually left out of the discussion, mainly because due to the closeness of the Landau-Plazcek-ratio to unity the Rayleigh-line is suppressed if compared to the longitudinal phonon peaks. However, technological progress in measurement allows to detect the Rayleigh-line in light-scattering experiments \cite{Nelson95a,Nelson95b}. In order to provide a proper theoretical setting for these experiments, we briefly discuss how the theory can be extended to include energy-fluctuations. The very formal presentation of the projection-operator formalism allows an easy and transparent modification of the equations.

%% Subsection: Definition and basic properties {{{2
\subsection{Definition and Basic Properties}\label{ :Temp_Def}
In this section, we will introduce the notion of a kinetic temperature whose fluctuations can be regarded as the variable being responsible for the Rayleigh-line in the spectra. According to \cite{Goetze89} that part of the kinetic energy fluctuations which is orthogonal to the mass-density fluctuation is a suitable quantity. Most of the definitions and formal properties can also be found in \cite{Goetze89}. Notice, that we decompose the total energy into $E = E^K + E^P$, i.e.,~kinetic and potential energy fluctuations respectively, and label the kinetic specific heat at constant volume by $c_V^0$. We define the temperature $\Theta$ by
\[\elabel{eq:def_temp}
c_v^0\Theta:={\mathcal Q}_\rho E^{K} = E^K - \frac{\bkt{E^K}{\rho}}{\bkt{\rho}{\rho}}\rho.
\]
This kinetic temperature is normalised to $c_v^0\bkt{\Theta(\q)}{\Theta(\q)}=k_BT^2$. The total energy is conserved and thus we have an additional local equation of continuity,
\[\elabel{eq:ener_cont}
\partial_t E = i\q\cdot\mb{J}^E.
\]
%% Using this definition we can infer 
%% \[
%% {\mathcal Q}E = {\mathcal Q}\left( {\mathcal Q}_\rho E^P + {\mathcal Q}_\rho E^K\right) = {\mathcal Q}E^{P}.
%% \]
Definition (\ref{eq:def_temp}) indicates, that the temperature is a $j=0$ object, i.e.,~it behaves like a scalar variable under rotation. Exploring the possible static correlation-functions we find, that within the chosen set of hydrodynamic variables in the long-wavelength limit temperature will only couple to the mass-density and the potential energy. Considering the microscopic form of these correlation-functions and the fact that the potential energy only depends on positions, but not on momenta, reveals that temperature is orthogonal to all variables that depend on positions only, in particular mass-density and potential energy. The projection-operator (\ref{eq:Project}) is therefore trivially extended by adding the normalised kinetic temperature. For static correlators one may also replace ${\mathcal Q}_\rho E^P =  {\mathcal Q}E^P$, where $\cal Q$ is a projection orthogonal to the extended hydrodynamic subspace, i.e.,~${\cal Q}\Theta = 0$.
The time-derivative of the kinetic temperature is, using the conservation of energy,
\begin{multline}
-ic_v^0\dot{\Theta}(q)=c_v^0{\mathcal L}\Theta(\q) = {\mc{LQ}}_\rho E^K \\= \q\cdot\mb{J}^E - \frac{\bkt{E}{\rho}}{\bkt{\rho}{\rho}}\q\cdot\mb{J} - {\mathcal LQ}E^P.
\end{multline}
This quantity does not define eigenvectors of the Liouville-operator for the eigenvalue $\q=0$ as the mass-current and the momentum-current do. We therefore must relax the assumption, that memory-kernels can be approximated by their $\q=0$ value. Indeed, we will actually find that the long-wavelength expansion of $\bkt{\dot{\Theta}}{R'(z)\dot{\Theta}}$ generates the heat-mode in the density-auto-correlator. 
%% In particular we find
%% \[
%% R'(t){\mc{LQ}}E^P = R'(t){\mc{QLQ}}E^P = -i\diff{t}R'(t)E^P,
%% \]
%% which after being Laplace-transformed reads
%% \[
%% R'{\mc{LQ}}E^P(z) = -\left(zR' + {\mathcal Q}\right)E^P(z).
%% \]
%% }}}2
%% Subsection: Correlation Functions involving Temperature {{{2
\subsection{Correlation-Functions involving Temperature}\label{ssec:Temp_Corr}

Due to the decoupling of the equations of motion (\ref{eq:MZW_2}) with respect to helicities, the only appearance of the temperature will be in the $\nu=0$ equations. By time-inversion parity the static correlation-functions $\bkt{\rho}{\dot{\Theta}} = -\bkt{\Theta}{\dot{\rho}} = 0$ as well as $\bkt{\Theta}{\dot{\Theta}}=0$. Using the equation of continuity for the mass-current (\ref{eq:EoC_curr})
\[\elabel{eq:C_temp_curr}
\bkt{J_0}{\dot{\Theta}}  = \frac{iq}{c_V^0}\bkt{p}{{\mathcal Q_\rho}E^K} + \order{q^2}.
\]
The correlation-functions involving the reduced evolution-operator $R'$ and the thus obtained transport-coefficients are computed in the following. Beginning with possible correlations to the mass-density we find
\[
\bkt{\dot{\rho}}{R'\dot{\Theta}} = -iq\bkt{J_0}{R'\dot{\Theta}} = -iq\bkt{J_0}{{\mathcal Q}R'\dot{\Theta}} = 0.
\]
Thus there are no correlations between mass-density and temperature. Next, we consider the correlations to the mass-current,
\[\begin{split}
\bkt{\dot{J}_0}{R'\dot{\Theta}} & = -iq \bkt{p}{R'\dot{\Theta}} + \order{q^3}\\
& = \frac{-q}{c_v^0}\bkt{p}{R'{\mathcal LQ}E^P} + \order{q^2}\\
& = \frac{q}{c_v^0}\left[z\bkt{p}{R'E^P} + \bkt{p}{{\mathcal Q}_{(\rho)}E^P}\right].
\end{split}
\]
The $\rho$ in brackets in the last line may be put or not due to the possible replacement ${\mathcal Q}_\rho E^P =  {\mathcal Q}E^P$. Using equations (\ref{eq:MZW_2}, \ref{eq:C_temp_curr}) we have a matrix element (cf.~\cite{FFL01})
\[\elabel{eq:dyn_tension}
\begin{split}
-i\bkt{\dot{J}_0}{\Theta} + \bkt{\dot{J}_0}{R'\dot{\Theta}} & = \frac{q}{c_v^0}\left[\bkt{p}{{\mathcal Q}_\rho E} + z \bkt{p}{R'E^P}\right]\\
& =: \frac{qmk_BT^2}{c_v^0}\beta(z)
\end{split}
\]
which defines a generalised dynamical tension coefficient $\beta(z)$. Under the reduced time evolution $\bkt{\dot{\Theta}}{R'\dot{\Theta}}$ is an autocorrelation-function. According to (\ref{eq:sel_00}) its expansion in powers of $\q$ reads
\[\elabel{eq:temp_temp_mat}
\frac{(c_V^0)^2}{k_BT^2}\bkt{\dot{\Theta}}{R'\dot{\Theta}} = z\tilde{c}_V(z)+ q^2\lambda(z) + \order{q^4}.
\]
The generalised specific heat $c_V(z)$ is defined $c_V(z):=c_V^0+\tilde{c}_V(z)$. The $\q^2$ terms of the reduced temperature-autocorrelation-function define the generalised thermal conductivity $\lambda(z)$. For the definitions and terminology of these coefficients see also \cite{Goetze89}.

We can now compose the matrix $\tilde{M}_0(z)$ and have therefore extended (\ref{eq:M_0}) to a model which incorporates energy-fluctuations. Temperature is simply added in the fifth row and column of the correlation matrix.
\begin{widetext}
\[\elabel{eq:M_0_energy}
\tilde{M}_{\nu=0}=\begin{pmatrix}
zS_{nr} & i\Omega_{nr} & 0 & 0 & 0\\
-i\Omega_{nr} & z\Omega_{nr} + \tilde{\Gamma}_{nr} & -iqC\tilde{\mu}_n & 0 & 0\\
0 & iqC\tilde{\mu}_r^T  & q^2\eta + zm^2v^2 & qm^2v^2 & \frac{qmk_BT^2}{c_v^0}\beta \\
0 & 0 & qm^2v^2 & z \frac{m^2v^2}{c^2} & 0\\
0 & 0 & \frac{qmk_BT^2}{c_v^0}\beta & 0 & \frac{k_BT^2}{(c_V^0)^2}\left(q^2\lambda + zc_v\right)
\end{pmatrix}.
\]
\end{widetext}
The solutions of the such modified equations of motion will only slightly differ from those which we have obtained in section \ref{sec:solution}. The autocorrelators of the orientational density and that of the mass-density are modified in the longitudinal-phonon-dissipation-kernel $K(z)$. One therefore has to replace
\[\elabel{eq:def_new_Gamma}
\frac{1}{mn}K(z)\mapsto\frac{1}{mn}K(z)-\frac{mT\beta(z)^2}{q^2\lambda(z)+zc_V(z)},
\]
introducing an additional wave-vector dependence in the damping $K(z)$, which will therefore be denoted by the symbol $K(z,q)$. The temperature-autocorrelation-function is 
\begin{multline}
-\mc{L}_{\Theta\Theta}(z) = \frac{k_BT^2}{q^2\lambda(z)+zc_V(z)} \\
+ \frac{zq^2T^2m^2v^2\beta(z)^2}{\left(q^2\lambda(z)+zc_V(z)\right)^2\left(c^2q^2-z^2-z\frac{q^2}{mn}K(z,q)\right)}.
\end{multline}
Here, $K(z,q)$ is the modified damping (\ref{eq:def_new_Gamma}) with the modification (\ref{eq:def_new_Gamma}). In the remainder of this section, we will discuss the question whether the modified phonon-prop\-a\-gator $\smash[t]{[c^2q^2-z^2-z\frac{q^2}{mn}K(z,q)]^{-1}}$ exhibits an additional hydrodynamic mode, which is visible in the light-scattering spectra. We are not able to answer this question for the general solutions with the full frequency-dependence of the memory-functions. However, in the limit of high temperatures we can use a Markov-approximation for the retarded correlators. The modification of the phonon-dissipation-kernel $K(z,q)$ only affects the polarised spectrum $I^{VV}$, cf.~equation (\ref{eq:I_VV_final}). We can therefore consider the autocorrelation of mass-density (\ref{eq:sol_dens}) which enters the expression for $I^{VV}$ using the modified phonon-dissipation-kernel $K(z,q)$ to find the heat-mode,
\[
\LT{\bkt{\rho(q)}{\rho(q,t)}}(z) = \frac{(mv)^2}{c^2}\frac{1}{z-\frac{q^2c^2}{z+\frac{q^2}{mn}K(z,q)}}.
\]
This correlation-function has the same functional form as the result for the density-auto-correlator that has been obtained in \cite{Goetze89} and \cite{FFL01}. Using Markov-approximations in the high-temperature limit 
\begin{eqnarray*}
K(z,q) \rightarrow i\eta^0,&
\lambda(z)\rightarrow i\lambda^0,\\
c_V(z)\rightarrow c_V+izc_V^{\prime\prime},&
\beta(z)\rightarrow\beta+iz\beta^{\prime\prime},
\end{eqnarray*}
with all constants being real and the corresponding limit of the memory-kernel matrix being positive definite, these authors have found three hydrodynamic modes, two sound waves and one heat-diffusion mode with poles at
\[
z_{\text{sound}} = \pm c_sq - iq^2\frac{\Gamma_l}{2} + \order{q^3},\qquad
z_{\text{heat}}= -iq^2D_T.
\]
The approximation for small $\q$ is due to \cite{Mountain66}. The adiabatic sound velocity is defined as $c_s^2:=c^2+mT\beta^2/c_V$ and the dissipation-constants are given by
\begin{eqnarray}
\Gamma_l &:=&\frac{\eta^0}{mn} + \frac{mT\beta^2}{c_V}\left[\frac{\lambda^0}{c_s^2c}+\frac{c_V^{\prime\prime}}{c_V}-2\frac{\beta^{\prime\prime}}{\beta}\right],\label{eq:def_gamma_l}\\
D_T &:=&\frac{c^2}{c_s^2}\frac{\lambda^0}{c_V}.\label{eq:def_D_T}
\end{eqnarray}
In contrast to the sole viscous and thermal damping of the sound waves by the first and the second term in (\ref{eq:def_gamma_l}) that is known from hydrodynamics \citep{Forster75}, $\Gamma_l$ has two additional contributions. These stem from the use of a kinetic temperature in our formulation. Since the total energy is conserved, there are contributions in the correlation-functions showing how kinetic energy is transformed into potential energy. This effect is irreversible and therefore displayed by two additional terms in the damping of the sound-waves $\Gamma_l$. The term with $c_V''$ describes the irreversible transfer of kinetic to potential energy at constant volume. Heating of the liquid due to mechanical work, i.e., the transformation of elastic energy into potential energy, is accounted for by the term with $\beta''$. 

%% }}}2

%% }}}1
%% Section: Derivation of previous results {{{1
\section{Comparing to Previous Results}\label{sec:Recovery}

Recently, a number of theoretical results for the problem of orientational dynamics and its application to light-scattering have been published \cite{Franosch03a, Franosch03b, FFL01, LaLe01}. The theoretical description given in these papers is comparable to the approach pursued in this work. However, the remarkable difference lies in the fact, that they directly use the decomposition of the dielectric tensor into its irreducible components $\varepsilon^{(j)}_\mu$ as basis-vectors for the hydrodynamic subspace, whereas we have preferred to project the dynamics onto the so-called tensorial densities $\rho_{\mu m}$ which by definition are closer related to the rotational dynamics of the molecules; see equations (\ref{eq:def_Q}, \ref{eq:def_rho_mn}) for the relation of the two quantities. The purpose of this section therefore is to relate our theory to the existing results by first presenting a theory, in the following called the {\em scalar} theory, which projects onto the variables $Q_\mu$ and then modify the equations of section \ref{sec:EOM} to yield exactly the same set of equations. It is obvious that the main concern is the modification of the memory-kernels and the question whether the newly derived kernels can still be interpreted as memory-kernels, i.e.,~as correlation-functions of a certain stochastic process. 

For the discussion of the latter we will define the notion of a correlation-function in an analytically tractable way as the characteristic function of a certain probability measure. Then properties of the Laplace-transform of such objects are used to prove that the functions which will appear in place of the memory-kernels of the scalar theory can indeed be interpreted as memory-kernels. The mathematical theory which we will have to employ is discussed in \cite{Akh65, Schultz03}. A brief overview regarding terminology can be found in Appendix \ref{app:moments}.

If one directly considers the irreducible components of the dielectric tensor as part of the basis of the hydrodynamic subspace in an application of the Mori-Zwanzig-technique, a suitable projection-operator will have the following form
\[\elabel{eq:def_P_scalar}
{\cal P}(\q) := \frac{\projb{Q_{\mu }}{Q_{\mu}}}{S} + \frac{\projb{\dot{Q}_{\mu}}{\dot{Q}_{\mu}}}{\Omega}
+ \frac{\proj{J_\mu}}{\bkt{J_0}{J_0}} + \frac{\proj{\rho}}{\bkt{\rho}{\rho}},
\]
where $\varepsilon^{(0)} \sim \rho$, $\varepsilon^{(2)}_\mu \sim Q_\mu$, and proper summation over $\mu$ is implied. The normalisation-factors are defined in a similar fashion as in section \ref{sec:hydro_var}. Accordingly, the matrices $\tilde{M}_\nu$ are derived by calculations following those of section \ref{sec:EOM}. For $\nu=0$ for instance it has the form
\[\elabel{eq:EoM_scalar_0}
\tilde{M}^{\text{sc}}_{\nu=0} = \begin{pmatrix}
zS & i\Omega & 0 & 0 \\
-i\Omega & z\Omega + \tilde{\Gamma} & -iqC_0\tilde{\mu} & 0\\
0 & iqC_0\tilde{\mu} & q^2\tilde{\eta} + zm^2v^2 & qm^2v^2\\
0 & 0 & qm^2v^2 & z \frac{m^2v^2}{c^2} 
\end{pmatrix},
\]
where the definition of the memory-kernels formally is the same as before. 

In the following discussion we will begin with giving results for the general case of dimensional reduction in Mori-Zwanzig-applications and then specialising to the case of orientational dynamics where appropriate.

%% Subsection: General Results {{{2
\subsection{Some General Results}

Consider the equation for a matrix of correlation-functions as given by Mori and Zwanzig in the form \cite{Forster75}
\[\elabel{eq:MZW_3}
C(z) = -S\left(z - \omega + M(z)\right)^{-1}.
\]
$C(z)$ represents the correlation-matrix, $S$ the normalisations, $\omega$ the libration frequencies, and $M$ the necessary memory-kernels. Without loss of generality we can assume that $M(z)$ is of order less than $\frac{1}{z}$. The general representation of analytic functions with positive imaginary part, as $M$ is such an object, has a term linear in $z$, one being constant and the remainder being of order less than $\frac{1}{z}$, see \cite{Akh65}. The linear and constant parts can be absorbed into $S$ and $\omega$. As can readily be seen equation (\ref{eq:MZW_3}) is equivalent to (\ref{eq:MZW_2}). The reduction to the scalar theory is done by multiplication with suitable matrices $\alpha$ from the left and the right. These produce the associated quadratic form for each block-matrix. The reduced correlation-matrix obeys the equation
\[\elabel{eq:C_red}
C_{\text{red}}(z) = \alpha^TC(z)\alpha = -\alpha^T S\left(z - \omega + M(z)\right)^{-1}\alpha,
\]
which can be cast into 
\[
C_{\text{red}}(z) = -S_{\text{red}}\left(z - \omega_{\text{red}} + M_{\text{red}}(z)\right)^{-1}.
\]
thus defining reduced quantities $S_{\text{red}}$, $\omega_{\text{red}}$, and $M_{\text{red}}$. As before we can assume $M_{\text{red}}$ to be of order less than $\frac{1}{z}$, because according to our discussion of the positivity properties of correlation-matrices, $M_{\text{red}}$ is an analytic function with positive imaginary part. 
We can use the high-frequency expansion of the correlator to gain more information. The zeroth moment of the representing measure of $M(z)$, cf.~Appendix \ref{app:moments}, will be denoted $m_0$.
\[\elabel{eq:exp_red}
C_{\text{red}}(z) \sim \alpha^T\left[\frac{-S}{z} - \frac{-S\omega}{z^2} - \frac{S}{z^3}\left(\omega^2 - m_0\right) - \dots \right]\alpha.
\]
The high-frequency expansion of $C_{\text{red}}$ with the reduced quantities is
\[
C_{\text{red}}(z) \sim \frac{-S_{\text{red}} }{z} - \frac{-S_{\text{red}}\omega_{\text{red}} }{z^2} - \frac{S_{\text{red}} }{z^3}\left(\omega_{\text{red}}^2 - m_{0,\text{red}}\right) - \dots 
\]
Comparing the coefficients we find 
\begin{eqnarray}
S_{\text{red}} &=&\alpha^TS\alpha\\
\Omega_{\text{red}} &=& \alpha^T S\omega\alpha\\
\omega_{\text{red}} &=& S_{\text{red}}^{-1}\Omega_{\text{red}}
\end{eqnarray}
and obtain an equation for the masses of $M_{\text{red}}$.
\[\elabel{eq:mk_red_mass}
m_{0,\text{red}} = S_{\text{red}}^{-1}\left[(Sm_0)_{\text{red}} - (\Omega\omega)_{\text{red}} + \Omega_{\text{red}}\omega_{\text{red}}\right],
\]
where $(Sm_0)_{\text{red}} =\alpha^\dagger Sm_0 \alpha$ and $(\Omega\omega)_{\text{red}}=\alpha^\dagger \Omega\omega\alpha$.
Thus, the measure which is representing the correlation-matrix $M_{\text{red}}$ has finite mass and renders the matrix a correlation-function in the sense of Appendix \ref{app:moments}. We note by the way, that the definitions of $S_{\text{red}}$ and $\omega_{\text{red}}$ are in accordance with the terminology and notation of \cite{Franosch03b}. That is normalisation and libration frequencies survive the reduction without any loss. The only changes will appear in the memory-kernels. From a physical point of view this is quite obvious. Each of the rotational variables has its own libration frequency. Contracting these generates beats. The frequency $\omega_{\text{red}}$ is the carrier, the remaining oscillations are put into a friction term, i.e.,~memory-kernel. The fact that this extension of the memory-functions does not alter their property of being the correlator of some stochastic process is shown by the above calculation. 

%% }}}2
%% Subsection: Number of Memory-kernels {{{2
\subsection{Number of Memory-kernels}

The remaining part to prove is that the number of memory-kernels in the reduced theory equals the number of memory-kernel matrices of the vector-valued theory. We do not bother with treating all helicities separately, but give the results for the most difficult case, $\nu=0$. Then the rest is obvious.
Consider the memory-kernel matrix of the vector-valued theory,
\[
M(z)=
\begin{pmatrix}
0 & 0 & 0 & 0 \\
0 & \Gamma_{mr} & -iqC\mu_m & 0\\
0 & \frac{iqC_0}{k_BTm}\tilde{\mu}_r^T  & \frac{q^2}{mn}\eta & 0\\
0 & 0 & 0 & 0
\end{pmatrix}.
\]
To compute the number and position of the non-vanishing elements of the reduced memory-kernel matrix we start by explicitly computing the respective masses $m_{0,\text{red}}$, defined in eq.~(\ref{eq:mk_red_mass}).
The matrices involved in its computation are the following.
\[
S = 
\begin{pmatrix}
\mb{S} & 0 & 0 & 0\\
0 & \mb{\Omega} & 0 & 0\\
0 & 0 & m^2v^2 & 0\\
0 & 0 & 0 & \frac{m^2v^2}{c^2}
\end{pmatrix}
\]
and
\[
\omega = 
\begin{pmatrix}
0 & -i\mb{\omega} & 0 & 0\\
-i\bm{1} & 0 & 0 & 0\\
0 & 0 & 0 & q\\
0 & 0 & qc^2 & 0
\end{pmatrix}
\]
where $\mb{\omega} = \mb{S}^{-1}\mb{\Omega}$ is used.
Thus
\[
\Omega = S\omega =
\begin{pmatrix}
0 & -i\mb{\Omega} & 0 & 0\\
-i\mb{\Omega} & 0 & 0 & 0\\
0 & 0 & 0 & (mv)^2q\\
0 & 0 & (mv)^2q & 0
\end{pmatrix}
\]
and hence
\[
\Omega\omega = 
\begin{pmatrix}
\mb{\Omega} & 0 & 0 & 0\\
0 & -\mb{\Omega\omega} & 0 & 0\\
0 & 0 & -(mvq)^2 & 0\\
0 & 0 & 0 & -(mvq)^2
\end{pmatrix}.
\]
The matrix $\Omega_{\text{red}}\omega_{\text{red}}$ has a form analogous to this. Hence we infer that the only non-vanishing element of the difference
$\Omega_{\text{red}}\omega_{\text{red}} - (\Omega\omega)_{\text{red}}$ is the $(2,2)$-element. $S^{-1}_{\text{red}}$ is diagonal, and will not change this fact. Since we know the form of $m_0$, we find that $m_{0,\text{red}}$ has only the elements $(2,2)$, $(2,3)$, $(3,2)$, and $(3,3)$ non-vanishing. 
It is a fact from function-theory, that the Stieltjes-transform of a measure has zeros on the diagonal, whenever the mass-matrix vanishes there \cite{Tsek00}. The imaginary part of the memory-kernel matrix $M_{\text{red}}$ is positive definite. For positive definite matrices $A$ we know $|A_{jk}|\le A^{1/2}_{jj}A^{1/2}_{kk}$. And therefore $(\Im M_{\text{red}})_{ij}=0$ for all $(i,j)$ in $\{(1,k), (k,1), (4,k), (k,4)\,|\, 1\le k \le 4\}$. However, the memory-kernels are analytic functions and thus the real part of these matrix-elements can only be a constant which by the assumptions on $M_{\text{red}}$ is zero. Hence the reduced memory-kernel matrix has the structure
\[
M_{\text{red}} = 
\begin{pmatrix}
0 & 0 & 0 & 0 \\
0 & A & B & 0 \\
0 & C & D & 0 \\
0 & 0 & 0 & 0
\end{pmatrix}.
\]
This means we have four memory-kernels at most! Now, we have to prove, that it is actually three distinct memory-kernels, in particular we show that $B$ is proportional to $C$. We can give an explicit expression for $M_{\text{red}}$ (also cf.~\cite{Akh65}~pp.~111.),
\[
M_{\text{red}}(z) = -z + \omega_{\text{red}} + S_{\text{red}}C_{\text{red}}^{-1}(z).
\]
The first term only affects the diagonals $A$ and $D$ and the matrix $\omega_{\text{red}}$ has no nonzero elements in the positions of interest. So we only have to care about the inverse of the reduced correlation matrix. The key of the proof are the specific symmetry properties of $C_{\text{red}}^{-1}$. The reduced correlation matrix has the following form
\[
C_{\text{red}} = 
\begin{pmatrix}
\LC{s}{s} & \LC{s}{\dot{s}} & \LC{s}{\nu} & \LC{s}{\rho}\\
\LC{\dot{s}}{s} & \LC{\dot{s}}{\dot{s}}  & \LC{\dot{s}}{\nu} & \LC{\dot{s}}{\rho}\\
\LC{\nu}{s} & \LC{\nu}{\dot{s}} & \LC{\nu}{\nu} & \LC{\nu}{\rho} \\
\LC{\rho}{s} & \LC{\rho}{\dot{s}} & \LC{\rho}{\nu} & \LC{\rho}{\rho}
\end{pmatrix}.
\]
The subscripts of $\cal L$ denote the various variables which are correlated in order to form the respective correlation-function. The subscript $s$ denotes the rotational degrees of freedom, $\dot{s}$ the associated currents, $\nu$ the mass-current, and $\rho$ the mass-density. We remember that for irreducible representations of the rotation group the interchange of variables in correlation-functions obeys equation (\ref{eq:sel_rule_dyn}).
Thus the parities under exchange of variables form the following matrix
\[
\begin{pmatrix}
+ & - & + & + \\
- & + & - & - \\
+ & - & + & + \\
+ & - & + & + 
\end{pmatrix}.
\]
Now, how do the symmetries transfer to particular elements of the inverse matrix? In general, the inverse $A^{-1}$ of a matrix $A$ has elements
\[
A^{-1}_{jk} = \frac{1}{\det A}(-)^{j+k}\hat{A}_{kj}.
\]
With $\hat{A}_{kj}$ we denoting the determinant of the matrix which is identical to $A$ except that the \th{k} row and the \th{j} column are missing. In particular, we are interested in the elements $A^{-1}_{23}$ and $A^{-1}_{32}$. 
Thus $\hat{A}_{23} = - \hat{A}_{32}$. This shows that the elements of interest in the inverse of the reduced correlation matrix are related to each other by real scalars. In more detail
\begin{eqnarray}
M_{\text{red}}(z)_{23} &=& \frac{1}{\Omega_{\text{red}}}C_{\text{red}}^{-1}(z)_{23}\\
M_{\text{red}}(z)_{32} &=& \frac{-1}{k_BTm}\Omega_{\text{red}}M_{\text{red}}(z)_{23}.
\end{eqnarray}
This shows the formal equivalence of the vector-valued theory and the scalar theory. To proceed from the first to the second one only has to replace the block-matrices by scalars. The memory-kernels retain their property of being correlation-functions. And due to positivity even the necessary condition on the memory-kernels \cite{Franosch03b} turns out to be the same.
%% }}}2

%% }}}1
%% Section: Conclusions {{{1
\section{Conclusions}\label{sec:concl}

In the present work, we have given a complete and comprehensive account of the long-wavelength rotational dynamics of a super-cooled molecular liquid and computed expressions for the light-scattering spectra of such systems. The theory is based on the concept of generalised hydrodynamics which views the transport-coefficients as frequency-dependent quantities, and an application of the projection-operator-technique of Mori and Zwanzig. The frequency-dependence of the transport-coefficients which appear as integral-kernels in the generalised Langevin equations allows both the description of the liquid in the high-temperature regime, where such dependence is negligible and the physical behaviour is that of a simple liquid, and the modelling of the solidification of the liquid near the glass-transition as it is known from mode-coupling theory. In contrast to previous approaches that only consider a decomposition of the dielectric tensor into its irreducible components, we model the complete rotational dynamics by using tensorial densities derived from Wigner-functions. The such obtained equations of motion can easily be solved using Laplace-transform methods and are used to obtain expressions for the light-scattering spectra both in the polarised and the depolarised configuration. We were able to reproduce the main features of the experimentally measured results, i.e., the Rytov-Dip at high temperatures and its vanishing in favour of propagating shear-waves after cooling. The importance of the translation-rotation coupling not only for the Rytov-Dip has been demonstrated by showing that the intensity in the VV configuration is not $\frac{4}{3}$ times that of the VH configuration in backscattering-geometry plus the isotropic contribution, thus contradicting the classical result for simple liquids \cite{BP00}. This difference, the so-called $VV$-dip, due to the translation-rotation coupling has recently been measured \cite{Cummins03b}.
The role of energy-fluctuations for the spectra has also been discussed. The theoretical model has been extended by also considering hydrodynamic modes of energy-fluctuations by similar techniques as used before. The appearance of an additional line in the scattering-spectrum due to heat, the Rayleigh-line, has also been shown.
Finally we were able to derive equations of motion from our more general theory which are formally equivalent to those obtained by only considering the irreducible components of the dielectric tensor itself in the projection-operator formalism. Using results from the theory of positive functions and the classical moment problem of probability theory we were able to characterise the Laplace-transforms of correlation-functions as those positive functions which satisfy a certain growth condition. This result enabled us to identify the functions which appeared in place of the memory-kernels in the simpler theories as correlation-functions, hence supporting their interpretation as memory-kernels as well.

%% }}}1

%% Appendices

\appendix
%% Appendix: Selection Rules {{{1
\section{Selection Rules}\label{app:sel_rule}
Due to the rotational invariance of the Hamiltonian function we can state a couple of selection rules. These will greatly simplify the evaluation of correlation-functions. We always assume that $\q$ is given parallel to the z-axis, i.e.,~$\q=q\bm{e}_0$. Most of these rules can be derived easily, for a more complete presentation see \cite{Fabbian00}.
Given two variables $A$ and $B$ lying in irreducible representations of rank $i$ and $j$ respectively it follows
\[\elabel{eq:sel_rule_gen}
\bkt{A_\mu^{(i)}(q)}{B_{\nu}^{(j)}(q)} = \delta_{\mu\nu}S^{ij}_\mu(q).
\]
Using rotation-matrices and playing around a little one finds
\[\elabel{eq:sel_rule_cc}
S^{ij}_{\mu}(q)^\ast = (-1)^{i+j}S^{ij}_{\mu}(q)
\]
and
\[\elabel{eq:sel_rule_neg}
S^{ij}_{\mu}(q) = (-1)^{i+j}\varepsilon^q_A\varepsilon^q_BS^{ij}_{-\mu}(q),
\]
where $\varepsilon^q$ denotes the parity under spatial reflection for each variable. In the special case $\q=0$ an even stronger selection rule applies
\[\elabel{eq:sel_rule_hel}
\bkt{A_\mu^{(i)}}{B_{\nu}^{(j)}} = \delta_{\mu\nu}\delta_{ij}S^{i}.
\]
If we expand the correlator (\ref{eq:sel_rule_gen}) in powers of $q$ the first non-vanishing term will be of order $|i-j|$. Two useful examples are
\[\elabel{eq:sel_00}
S^{00}_{0}(q) = S^{00} + \order{q^2}
\]
and
\[\elabel{eq:sel_02}
S^{02}_{\mu}(q) = \order{q^2}.
\]
The Hamiltonian function is invariant with respect to time inversion. Hence we can assign time-inversion-parities $\varepsilon^t_A$ to the dynamical variables. And therefore we find for the behaviour of dynamical correlators under exchange of variables
\[\elabel{eq:sel_rule_dyn}
\bkt{A_\mu^{(i)}(q,t)}{B_{\mu}^{(j)}(q)} = (-1)^{i+j}\varepsilon^t_A\varepsilon^t_B\bkt{B_\mu^{(j)}(q,t)}{A_{\mu}^{(i)}(q)}.
\]
Time inversion parities and the tensorial rank of dynamical variables are preserved under time evolution.

%% }}}1
%% Appendix: Normalisation {{{1
\section{Normalisation}\label{app:norm}
In this Appendix we show the invertibility of the normalisation matrices $S_{mn}$ and $\Omega_{mn}$ as they have been defined in section \ref{sec:hydro_var}. Due to the simplicity of the proof we can deal with general $j$ here. The case needed in the present paper is $j=2$ only. The matrix $S_{mn}$ as defined in equation (\ref{eq:def_norm}) is a real and symmetric matrix on a space $\mathbb{C}^{2j+1}$ by property (\ref{eq:sel_rule_cc}). Thus one may diagonalise it. What is left to prove its invertibility is, that none of the eigenvalues vanishes.
A sufficient condition is, that the associated linear operator is positive definite. Taking an arbitrary vector $\lambda \in \mathbb{C}^{2j+1}$ we compute the scalar product 
\[
\sum_{mn} \lambda_{m}^{\ast}S_{mn}\lambda_{n}
=\erw{\sum_{mn}\lambda_{m}^{\ast}\rho_{m}^{\ast}\lambda_{n}\rho_n}  
=\erw{|\sum_{m} \lambda_m\rho_m|^2}\ge 0.
\]
Since the Kubo-scalar product is non-degenerate we only need to look whether there is a certain $\lambda$ to make the sum vanish.
The proof goes indirectly:
Let $0 \neq \lambda \in \mathbb{C}^{2j+1}$ such that $\sum_{m}\lambda_m\rho_m = 0$.  Recalling the definition of $\rho_{\mu m}$ (\ref{eq:def_rho_mn}) this condition is equivalent to
\[
\sum_{\alpha=1}^{N} e^{i\q\bm{R}_\alpha}\sum_{m}{\cal D}^{(j)}_{\mu m}(\Omega_\alpha)\lambda_m = 0.
\]
We assume that all the $\bm{R}_\alpha$ are different and since the exponentials are orthogonal we can project onto the contribution of one single molecule $\alpha$,
\[
\sum_{m}{\cal D}^{(j)}_{\mu m}(\Omega_\alpha)\lambda_m = 0.
\]
Multiplying from the right with ${\cal D}^{(j)}_{-\mu, -n}(\Omega_\alpha)^{\ast}$ and integrating over the group we can exploit the orthogonality between the representations \cite{Wigner59} and find $\lambda_n = 0$. Since $n$ was arbitrary $\lambda = 0$ and the proof is given.

To prove that the normalisation matrix of the orientational current is invertible, is more complicated. Using angular momenta instead of Eulerian angles we formulate the orientational kinetic energy as
\[\elabel{eq:H_rot_kin}
T^{\text{rot}} = \sum_\alpha \sum_{i=1}^{3}\frac{J_{\alpha,i}^2}{2I_i}.
\]
Some algebra yields
\[\elabel{eq:Omega_mn}
\Omega_{mn} = \frac{2j+1}{8\pi^2}k_BT \sum_k \int \D\Omega\,{\cal D}^{(j)*}_{\mu m}(\Omega)\frac{L_k^2}{I_k}{\cal D}^{(j)}_{\mu n}(\Omega).
\]
This is nothing but the quantum mechanical asymmetric rigid rotor whose solution can be taken from \cite{LL3_66}.
Like the matrix $S_{mn}$, the normalisation-matrix of the orientational currents $\Omega_{mn}$ is real and symmetric. Hence it is diagonalisable. We only need to show its positive definiteness. All involved operators appear quadratically and because the $L_i$ are self-adjoint for $i=x,y,z$, their eigenvalues are real and the eigenvalues of their squares are nonnegative. Consider that contribution to the matrix which is proportional to $L_z^2$. This operator is already diagonal and its matrix-elements are
\[
(L_z^2)_{mm} \propto m^2\quad\text{with}\quad m\in[-j;j] \cap \Z.
\]
Let $\lambda_m\in \C^{2j+1}$ be the eigenvectors of $L_z$. The operator $L_z^2$ has a non-vanishing kernel, namely $K:=\lin\{\lambda_0\}$. Obviously, using the casimir operator $\mb{L}^2$, the identity $\mb{L}^2 - L_z^2 = L_x^2 + L_y^2$ holds and $\lambda^\ast_0(L_x^2+L_y^2)_{00}\lambda_0 = \lambda^*_0(\mb{L}^2)_{00}\lambda_0 > 0$.
When $K$ is the kernel of $\Omega_{mn}$, we can either have $\lambda_0^\ast (L_x^2)_{00}\lambda_0=0$ or the respective quadratic form of $L_y^2$ vanishing, but not both, because their sum is positive. The only chance to find a zero eigenvalue would have been to have a common kernel of all three operators. The above relation shows that this cannot be accomplished and hence $\Omega_{mn}$ is positively definite and therefore invertible.
%% }}}1
%% Appendix: Positive Functions {{{1
\section{Positive Functions}\label{app:moments}
To accomplish the reduction of the presented theory of light-scattering in super-cooled liquids to a scalar theory comparable to \cite{Franosch03a, Franosch03b}, some mathematical definitions and results of the theory of positive functions are needed.
We define a correlation-function to be the characteristic function of a probability measure \cite{Feller68, Shiryaev96, Goetze95}. 
\[\elabel{def:char_function}
\phi(t) := \int\limits_{\Omega}e^{-i\lambda t}\D\mu(\lambda),
\]
where without loss of generality $\int \D\mu(\lambda)=1$. One can easily compute its Laplace-transform
\[\elabel{eq:stieltjes}
\LT{\phi}(z) = \int\frac{1}{\lambda-z}\,\D\mu(\lambda).
\]
This means that the Laplace-transform of a correlation-function is the Stieltjes-transform of a probability measure. In particular, it is an analytic function mapping the complex upper half-plane into itself.
In connection with the solution of the classical moment problem \cite{Akh65}, Hamburger and Nevanlinna have proved that for an analytic function $f$ mapping the upper half-plane into itself being the Stieltjes-transform of a finite measure it is necessary and sufficient that 
\[\elabel{eq:crit_corr_fn}
\sup_{y\ge 1}|yf(iy)| <\infty.
\]
The moments $s_k$ of the measure $\mu(\lambda)$ are given by the coefficients of the large $z$ expansion of $f$,
\[
f \sim -\frac{s_0}{z} - \frac{s_1}{z^2} - \frac{s_2}{z^3} - \dots,
\]
Often the moments of the representing measures are called the moments of $f$. Thus an analytic function mapping the upper half plane into itself is regarded the Laplace-transform of a correlation-function if and only if the above criterion (\ref{eq:crit_corr_fn}) on the growth of $f$ holds true.
%% }}}1
%% Backmatter
\providecommand{\SortNoop}[1]{}

\end{document}